\documentclass[11pt]{article}

\pdfoutput=1

\usepackage[all,arc,curve,frame]{xy}

\usepackage{bookman} 

\usepackage[english]{babel}
\usepackage[latin1]{inputenc}

\usepackage{amsmath}
\usepackage{amssymb}
\usepackage{theorem}

\usepackage{listings}
\usepackage{alltt}

\usepackage{epsfig}

\usepackage{color}

\usepackage{graphicx}

\usepackage{comment}

\newcommand{\writes}{\mapsto}

\renewcommand{\implies}{\Rightarrow}
\newcommand{\isImplied}{\Leftarrow}

\newcommand{\lOr}{\,\vee\,}
\newcommand{\lAnd}{\,\wedge\,}

\newcommand{\union}{\,\cup\,}

\newcommand{\naturals}{\mathbb{N}}

\newcommand{\related}{\rightsquigarrow}

\newtheorem{definition}{Definition}

\newcommand{\fin}{$\Box$}

\newcommand{\such}{~{\rule[-0.2em]{0.25ex}{1em}}}

\title{Formal Definitions of\\ Memory Consistency Models}
\author{Jordi Bataller Mascarell}
\date{\today}

\begin{document}


\maketitle

\begin{abstract}

Shared  Memory  is  a  mechanism  that  allows  several  processes  to
communicate with each  other by accessing --writing or  reading-- a set
of variables that they have in common.

A Consistency Model defines how each process observes the state of the
Memory, according to  the accesses performed by it and  by the rest of
the processes  in the  system. Therefore, it  determines what  value a
read returns when a given process  issues it.  This implies that there
must  be an  agreement  among  all, or  among  processes in  different
subsets, on  the order  in which  all or a  subset of  the accesses
happened.

It is  clear that a  higher quantity  of accesses or  proceses taking
part  in the  agreement  makes  it possibly  harder  or  slower to  be
achieved.  This is the main reason for which
a number of Consistency Models for Shared Memory
have been introduced.

This paper is a handy summary of 
\cite{Bataller:Synchronized:1997} and 
\cite{Bernabeu:Formalizing:1994} where consistency models
(Sequential,  Causal,   PRAM,  Cache,  Processors,   Slow),  including
synchronized ones (Weak, Release, Entry), were formally defined.  This
provides a  better understanding of those  models and a way  to reason
and compare them through a concise notation.

There  are  many  papers  on  this  subject  in  the  literature  such
as \cite{Steinke:Unified:2004} with which this work shares some concepts.


\end{abstract}



%
\section{Fundamentals}


A memory is a system that accepts two operations: {\em write} and {\em
  read}.  These operations  can be issued by any process  from the set
${\cal P}$; and are related to one  of the variables in the set ${\cal
  V}$.  As  usual, {\em write}  sets a  new value for  a variable,
whereas {\em read} returns the  value associated with a variable.  For
sake  of simplicity  and without  loss of  generality, we  assume that
${\cal P} \subset  \naturals$, all variables are  of type $\naturals$,
and that  writes are {\em  uni-valued} (a  given value may  be written
only once).

We designate write and read operations with the following notation.


\begin{definition}{Write and Read}

\begin{itemize}
\item $w(i,v,a)$ to
  denote that process $i \in \cal P$ writes the value $a \in \naturals$
  to the variable $v \in \cal V$.
\item $r(i,v,a)$ to
  denote that process $i \in \cal P$ reads $a \in \naturals$
  from the variable $v \in \cal V$.
\end{itemize}

\fin
\end{definition}

\begin{definition}{Sequence-Execution}

A  sequence-execution of a memory  system is a sequence of write
and read operations.  For a given execution $\alpha$, its associated total order $<^\alpha$ 
is trivially defined as

$$
o_1  <^\alpha  o_2   \equiv  \alpha  =  \alpha'   o_1  \alpha''  o_2
\alpha'''
$$
\fin
\end{definition}

We can apply a condition $c$ to filter a sequence-execution $\alpha$
as well as an order $<^\alpha$.

\begin{definition}{Filtering}

\begin{itemize}

  \item $\alpha | c = $

    \[
    \begin{cases}
      \epsilon & \isImplied \alpha = \epsilon \\
      \begin{cases}
         o\ (\alpha_1|c) & \isImplied c(o) \\
         \alpha_1|c & \isImplied \neg c(o)
      \end{cases} & \isImplied \alpha = o\ \alpha_1
    \end{cases}
    \]

    %
    
    


\item $
   o_1 (<^\alpha|c) o_2 \ \equiv $
  $o_1 <^\alpha o_2  \lAnd c(o_1) \lAnd c(o_2)
  $

%

\end{itemize}

\fin
\end{definition}

Common filters include
\begin{itemize}

  \item $( \Box | i: \cal P)$, $\Box$ (sequence or order) restricted to actions process $i$.
  \item $( \Box | v: \cal V)$, $\Box$ restricted to actions on variable $v$.
  \item $( \Box | w )$, $\Box$ restricted to write actions.
  \item $( \Box | r )$, $\Box$ restricted to read actions.
  \item $( \Box | w(i,\cdot,\cdot))$, $\Box$ restricted to write actions by process i.
  \item  $( \Box  | w(\cdot,  v, \cdot))$,  $\Box$ restricted  to write
    actions on variable $v$.

  \item $( \Box | (a, b) ) \equiv (\Box | a) \union (\Box | b)$

\end{itemize}

We only consider  {\em valid} executions.  An execution  is valid when
every read gets its value from a {\em previous} write.

\begin{definition}{Valid Execution}
\label{def:ValidExecution}

$\alpha$ is a valid execution $\equiv$

\begin{itemize}

\item 
  $r(\cdot, v, a) \in \alpha \implies \alpha  =  \alpha_1\, w(\cdot,v,a)\,  \alpha_2\,  r(\cdot, v,  a)\, \alpha_3 $


\item
$w(i, v, a) \in \alpha \equiv \alpha | w(\cdot, v, a) = w(i, v, a)$
  (writes are uni-valued)

\end{itemize}
\fin
\end{definition}

Valid executions  are not  meant to  capture ``real  time'' accurately
(time is relative in a distributed system).   We are only
forbidding that a read gets a value random value or either a value from
the future.

A valid execution $\alpha$ defines two relationships: writes-to and
process order.

\begin{definition}{Writes-to and Process Order}

  For a valid execution $\alpha$:

\begin{itemize}
\item {\em writes-to} relates
a read and the write which set the value:

$o_1 \writes_\alpha o_2 \equiv 
o_1 = w(\cdot, v, a) \in \alpha
\lAnd
o_2 = r(\cdot, v, a) \in \alpha
$

\item {\em process order} relates all the actions by the same process:

  $o_1(i,\cdot,\cdot) <_{PO}^\alpha o_2(j,\cdot,\cdot) \equiv
  i=j \lAnd 
  o_1 <^\alpha o_2$

  Alternate definition:

  $a <_{PO}^\alpha b  \equiv (\exists i \in  P \such : (\alpha  | i) =
  \alpha_1\, a\, \alpha_2\, b\, \alpha_3)$

\end{itemize}

\fin
\end{definition}

The  transitive  closure  of  the  writes-to  and  the  program  order
relations defines  a {\em partial order}  over the actions of  a valid
execution $\alpha$ called the {\em causal relation}.

\begin{definition}{Causal Relation}

$$<_{CR}^\alpha \equiv (\writes_\alpha \cup <_{PO}^\alpha)^*$$
\fin
\end{definition}

A valid execution is {\em consistent} if
it contains no read fetching an
overwritten value.

\begin{definition}{Consistent Execution}

$\alpha$ is consistent $\equiv$

\[
r(i,v,a) \in \alpha \implies \\
    \begin{cases}
      \alpha  = \alpha_1\  w(j,v,a)\  \alpha_2\  r(i,v,a)\ \alpha_3  ~ \lAnd ~ w(j,v,a)\writes r(i,v,a) )\\ \lAnd \\
      \alpha_2|w(\cdot,v,\cdot) = \epsilon
    \end{cases}
\]
%

\fin
\end{definition}

A key concept used for defining models is {\em linearization}
which captures the idea of extending  a partial order to a total
one while respecting consistency.

\begin{definition}{Linearizability}

  A relation $<_?^\alpha$
  is consistently linearizable  (it has a consistent linear
  extension, linearizable for short) $\equiv$ 

$(\exists \beta :\textrm{\ sequence of  the actions in\ } \alpha \such : 
  \beta \textrm{\ is consistent} \lAnd <_?^\alpha \subseteq <^\beta$)

\fin
\end{definition}

Because $\beta$, the sequence proving  the linearizability of a given
$<_?^\alpha$,  is   consistent,  we  have   $\writes_\alpha  \subseteq
<^\beta$.   And   because  $<^\beta$   is  a  total   order  including
$<_?^\alpha$   the  closure   of  $<_?^\alpha$   and  $\writes_\alpha$
satisfies

$$(<_?^\alpha \union \writes_\alpha)^* \subseteq <^\beta$$

and  it   is,  obviously,  acyclic;   as  well  as   $<_?^\alpha$  and
$\writes_\alpha$ are.

In order to show that a sequence, $\beta$, proves that $<_?^\alpha$ is linearizable, is
enough for $\beta$ to include all actions in 
$<_?^\alpha$ plus  the writes not  yet included and necessary  for any
read
in  $<_?^\alpha$. The  rest of  actions of  $\alpha$ can  be trivially
added  to the  end of  $\beta$, only  having to  obey $\writes_\alpha$
since they are not in $<_?^\alpha$.

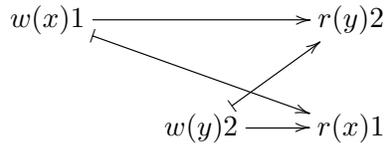
\begin{figure}[htbp]
\centerline{
\xymatrix{
  w(x)1 \ar[rr] \ar@{|->}[drr] & ~ & r(y)2 \\
  ~ & w(y)2 \ar[r] \ar@{|->}[ur] & r(x)1 
} 
} 
\caption{A diagram-execution.}
\label{fig:first:diagram}
\end{figure}

It's  much easier  to  understand  and to  reason  on executions  when
depicted  in  a {\em  diagram-execution},  where  process actions  are
visually separated and the causal  relation is explicit.  Notation can
also  be  simplified.  As  an introduction  of  concepts,  consider  the
diagram-execution in figure \ref{fig:first:diagram}.

Time increases from  left to right in the  diagram.  Horizontal arrows
express  process order.   Diagonal arrows  denote writes-to  order.  By
definition of process order and by the condition that reads always get
values previously written, valid diagrams may not have arrows pointing
left.  Aside  from this, diagrams  don't reflect (unless  stated) when
events happened in real time.  Hence, the fact that $w(2,y,2)$ is
to the right of
$w(1,x,1)$ doesn't imply that $w(2,y,2)$ happened necessarily later.

Clearly, a diagram-execution defines a set of
sequence-executions: those ones respecting process order and writes-to
as expressed by the diagram-execution.


%



\section{Consistency Models}

\subsection{Sequential Consistency}

The first memory  model we introduce is  {\em sequential consistency}
\cite{Lamport:How:1979}.  Compared with  the rest  of models  we
shall discuss, this one corresponds to the common understanding on how
a memory behaves in absence of a global clock.

A sequentially  consistent memory provides  a total ordering  of writes
(all processes agree on the order in which memory accesses happened)
and ensures that every read gets  always the last value written to its
variable.

\begin{definition}{Sequential Consistency}

$\alpha$ is an execution by a Sequential memory $\equiv$

  $$<^\alpha_{PO}\ \textrm{is consistently linearizable.} $$

\fin
\end{definition}

Note the implications of this definition.
If $\beta$ is the required consistent linear extension
of $\alpha$, then
$ <^\alpha_{CR} \subseteq <^\beta$
(and $<^\alpha_{PO} = <^\beta_{PO}$, $\writes_\alpha = \writes_\beta$).

\begin{figure}[htbp]
\centerline{
\xymatrix{
  w(x)1 \ar[rr] \ar@{|->}@/^1pc/[rr] \ar@{|->}[drr] & ~ & r(x)1 \\
  ~ & w(x)2 \ar[r] & r(x)1 
} 
} 
\caption{Sequential execution.}
\label{fig:sequential}
\end{figure}
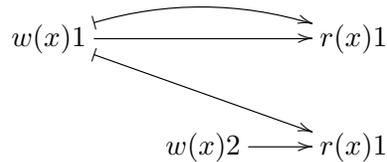

As a first example, note that the execution 
in   figure \ref{fig:sequential} is sequential. This
consistent linear extension 
$w(2,x,2)\ w(1,x,1)\ r(1,x,1)\ r(2,x,1)$ is the proof.

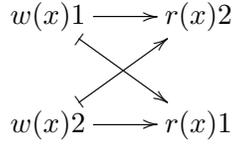
\begin{figure}[htbp]
\centerline{
\xymatrix{
  w(x)1 \ar[r] \ar@{|->}[dr] & r(x)2 \\
  w(x)2 \ar[r] \ar@{|->}[ur] & r(x)1 
} 
} 
\caption{Non-sequential execution.}
\label{fig:nonSequential}
\end{figure}

However, the execution
in figure \ref{fig:nonSequential} is not sequential.

Works  like \cite{Lipton:PRAM:1988}  suggest  that,  in a  distributed
system, a sequential memory can't be simulated with waiting-free write
and read  operations: the process  issuing an operation must  wait for
the response  from at least one  different process in order  to ensure
the common view of memory accesses.   This fact led to the proposal of
memory models with fewer consistency requirements.

\subsection{Causal Consistency}

The {\em causal consistency} memory model
\cite{Ahamad:Causal:1991}
allows two processes
to disagree on the order of writes only in case they are causally unrelated.

Causal dependencies  are very  easy to identify  on diagram-executions
because two  operations are causally related  if there is a  path from
one of them to the other.

\begin{definition}{Causal Consistency}

$\alpha$ is an execution by a Causal memory $\equiv$

$$(\forall i \in {\cal P} \such : <^\alpha_{CR} |(i,w) \ \textrm{is
consistently linearizable} )$$

\fin
\end{definition}

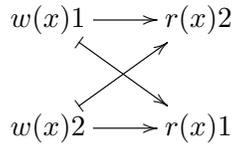
\begin{figure}[htbp]
\centerline{
\xymatrix{
  w(x)1 \ar[r] \ar@{|->}[dr] & r(x)2 \\
  w(x)2 \ar[r] \ar@{|->}[ur] & r(x)1 
} 
} 
\caption{Causal execution.}
\label{fig:causal}
\end{figure}

The execution $\alpha$ in figure \ref{fig:causal}
is causally consistent because

$<^\alpha_{CR} |(1,w) =$ 
\fbox{
\xymatrix{
  w(x)1 \ar[r] & r(x)2 \\
  w(x)2 \ar@{|->}[ur] & ~
} 
} 

and
$<^\alpha_{CR} |(2,w) =$
\fbox{
\xymatrix{
  w(x)1 \ar@{|->}[dr] & ~ \\
  w(x)2 \ar[r] & r(x)1 
} 
} 

have consistent linear extensions.
This extensions, expressing each process's point of view, don't agree
on the order of $w(1,x,1)$ and $w(2,x,2)$. This is acceptable
as these writes are causally unrelated.

\subsection{PRAM Consistency}

The Pipelined RAM memory model \cite{Lipton:PRAM:1988} further relaxes
requirements.  It allows  two processes  to disagree  on the  order of
writes if they are issued by different processes.

\begin{definition}{PRAM Consistency}

$\alpha$ is an execution by a PRAM memory $\equiv$

$$(\forall i \in {\cal P} \such : <^\alpha_{PO} |(i,w)
\ \textrm{is consistently linearizable} )$$

\fin
\end{definition}

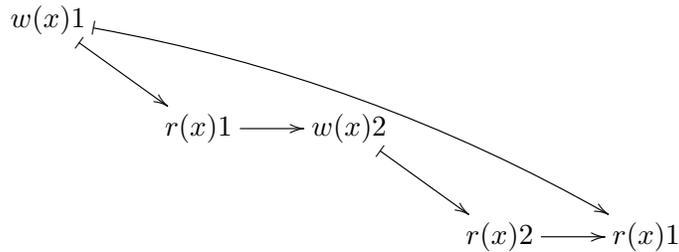
\begin{figure}[htbp]
\centerline{
\xymatrix{
w(x)1 \ar@{|->}[dr] \ar@{|->}@/^1pc/[ddrrrr] & ~ & ~ & ~ & ~ \\
~ & r(x)1 \ar[r] & w(x)2 \ar@{|->}[dr]& ~ & ~ \\
~ & ~ & ~ & r(x)2 \ar[r] & r(x)1 \\
} 
} 
\caption{PRAM execution.}
\label{fig:pram}
\end{figure}

The execution in figure \ref{fig:pram}
is PRAM consistent because

$<^\alpha_{PO} |(1,w)=$
\fbox{
\xymatrix{
w(x)1  & ~ & ~  \\
~ & ~ & w(x)2  \\
} 
} 

$<^\alpha_{PO} |(2,w)=$
\fbox{
\xymatrix{
w(x)1 \ar@{|->}[dr] & ~ & ~ \\
~ & r(x)1 \ar[r] & w(x)2 \\
} 
} 

$<^\alpha_{PO} |(3,w)=$
\fbox{
\xymatrix{
w(x)1 \ar@{|->}@/^1pc/[ddrrrr] & ~ & ~ & ~ & ~ \\
~ & ~ & w(x)2 \ar@{|->}[dr]& ~ & ~ \\
~ & ~ & ~ & r(x)2 \ar[r] & r(x)1 \\
} 
} 

have consistent linear extensions. However, it
is not causally consistent because

$<^\alpha_{CR} |(3,w)=$
\fbox{
\xymatrix{
w(x)1 \ar@{~>}[drr] \ar@{|->}@/^1pc/[ddrrrr] & ~ & ~ & ~ & ~ \\
~ & ~ & w(x)2 \ar@{|->}[dr]& ~ & ~ \\
~ & ~ & ~ & r(x)2 \ar[r] & r(x)1 \\
} 
} 

has no consistent linear extension.

\subsection{Cache Consistency}

This model, \cite{Goodman:Cache:1989}, focus on the consistency of each
  variables separately.   All processes  must agree  on the  order of
  accesses to the  same variable, but they are allowed  to disagree on
  accesses to different variables.

\begin{definition}{Cache Consistency}

$\alpha$ is an execution by a Cache memory $\equiv$

$$(\forall v \in {\cal V} \such : <^\alpha_{PO} |v
\ \textrm{is consistently linearizable} )$$

\fin
\end{definition}

\begin{figure}[htbp]
\centerline{
\xymatrix{
 w(x)1 \ar@/_2pc/@{|->}[drrrr] \ar[r] & w(x)2 \ar[r] & w(y)3 \ar@{|->}[dr] & ~ & ~\\
  ~ & ~ & ~ & r(y)3 \ar[r] & r(x)1 \\
} 
} 
\caption{Cache execution.}
\label{fig:cache}
\end{figure}
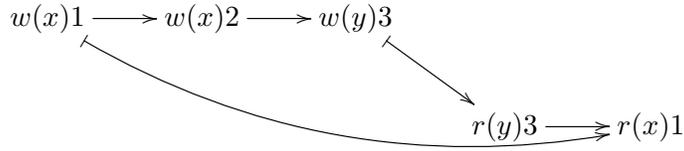

The execution  $\alpha$ in figure \ref{fig:cache}  is Cache consistent
because

$<^\alpha_{PO}|x =$
\fbox{
\xymatrix{
 w(x)1 \ar@/_2pc/@{|->}[drrrr] \ar[r] & w(x)2 & ~ & ~ & ~\\
  ~ & ~ & ~ & ~ & r(x)1 \\
} 
} 

and
$<^\alpha_{PO}|y =$
\fbox{
\xymatrix{
 ~ & ~ & w(y)3 \ar@{|->}[dr] & ~ & ~\\
  ~ & ~ & ~ & r(y)3 & ~ \\
} 
} 

have consistent linear extensions.
However, it is not PRAM nor causally consistent because

$<^\alpha_{PO}|(2,w) = <^\alpha_{CR}|(2,w) = $

\fbox{
\xymatrix{
 w(x)1 \ar@/_2pc/@{|->}[drrrr] \ar[r] & w(x)2 \ar[r] & w(y)3 \ar@{|->}[dr] & ~ & ~\\
  ~ & ~ & ~ & r(y)3 \ar[r] & r(x)1 \\
} 
} 

does not have a consistent linear extension.


\subsection{Processor Consistency}

This model, also defined in \cite{Goodman:Cache:1989}, 
could be viewed as the intersection
of the PRAM and Cache consistency models.
But, actually, it is a little stronger than just this.

\begin{definition}{Processor Consistency}

$\alpha$ is an execution by a PROC memory $\equiv$

\begin{multline*}
  (\forall i, j \in {\cal P} \such : 
  <^\alpha_{PO}|(i,w) \textrm{\ and \ }<^\alpha_{PO}|(j,w) \\
  \textrm{\ have, respectively, consistent linear extensions\ }
  \beta_i \textrm{\ and \ } \beta_j \\
  \lAnd 
(\forall x \in {\cal V} \such : \beta_i|w(\cdot,x,\cdot)
=\beta_j|(w(\cdot,x,\cdot) ) )
\end{multline*}

\fin
\end{definition}

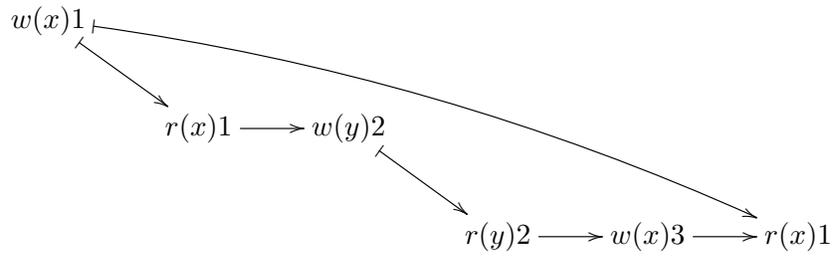
\begin{figure}[htbp]
\centerline{
\xymatrix{
w(x)1 \ar@{|->}[dr] \ar@{|->}@/^1pc/[ddrrrrr] & ~ & ~ & ~ & ~ & ~ \\
~ & r(x)1 \ar[r] & w(y)2 \ar@{|->}[dr]& ~ & ~ & ~ \\
~ & ~ & ~ & r(y)2 \ar[r] & w(x)3 \ar[r] & r(x)1 \\
} 
} 
\caption{Processor execution.}
\label{fig:processor}
\end{figure}

The execution $\alpha$ in figure \ref{fig:processor}
is processor consistent because

$<^\alpha_{PO}|(1,w) =$
\fbox{
\xymatrix{
w(x)1 & ~ & ~ & ~ & ~ \\
~ & ~ & w(y)2 & ~ & ~ \\
~ & ~ & ~ & ~ & w(x)3 \\
} 
} 

$<^\alpha_{PO}|(2,w) =$
\fbox{
\xymatrix{
w(x)1 \ar@{|->}[dr] & ~ & ~ & ~ & ~ & ~ \\
~ & r(x)1 \ar[r] & w(y)2 & ~ & ~ & ~ \\
~ & ~ & ~ & ~ & w(x)3 & ~ \\
} 
} 

$<^\alpha_{PO}|(3,w) =$
\fbox{
\xymatrix{
w(x)1 \ar@{|->}@/^1pc/[ddrrrrr] & ~ & ~ & ~ & ~ & ~ \\
~ & ~ & w(y)2 \ar@{|->}[dr]& ~ & ~ & ~ \\
~ & ~ & ~ & r(y)2 \ar[r] & w(x)3 \ar[r] & r(x)1 \\
} 
} 

have the following consistent linear extensions:

$\beta_1 = w(3,x,3)\ w(1,x,1)\ w(2,y,2)$

$\beta_2 = w(3,x,3)\ w(1,x,1)\ r(2,x,1)\ w(2,y,2)$

$\beta_3 = w(2,y,2)\ r(3,y,2)\ w(3,x,3)\ w(1,x,1)\ r(3,x,1)$

which, additionally, satisfy
$\beta_1|w(\cdot,x,\cdot) =
\beta_2|w(\cdot,x,\cdot) =
\beta_3|w(\cdot,x,\cdot)$ 
and
$\beta_1|w(\cdot,y,\cdot) =
\beta_2|w(\cdot,y,\cdot) =
\beta_3|w(\cdot,y,\cdot)$ 

However, it is clear that $\alpha$ is not causally consistent.

\begin{figure}[htbp]
\centerline{
\xymatrix{
w(x)1 \ar[r] \ar@{|->}@/_3pc/[ddrrrrr] & w(y)2 \ar@{|->}[dr] & ~ & ~ & ~ & ~ \\
~ & ~ & r(y)2 \ar[r] & w(x)3 \ar@{|->}[dr] & ~ & ~ \\
~ & ~ & ~ & ~ & r(x)3 \ar[r] & r(x)1 \\
} 
} 
\caption{PRAM and Cache execution.}
\label{fig:pramAndCache}
\end{figure}
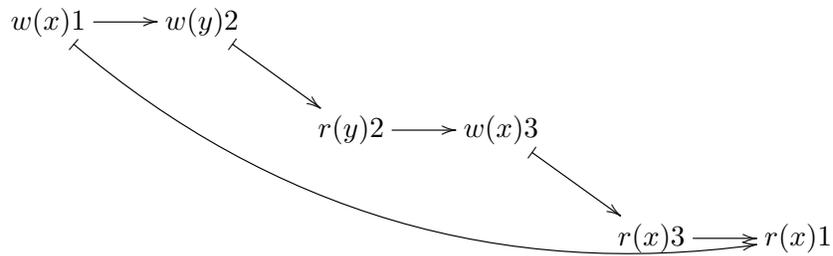

It's worth to mention that there  are executions that, being PRAM and
Cache consistent executions, are not Processor consistent.
For example, consider the execution $\alpha$ in figure \ref{fig:pramAndCache}.

$\alpha$ is Cache consistent because

$<^\alpha_{PO}|x =$
\fbox{
\xymatrix{
w(x)1 \ar@{|->}@/_3pc/[ddrrrrr] & ~ & ~ & ~ & ~ & ~ \\
~ & ~ & ~ & w(x)3 \ar@{|->}[dr] & ~ & ~ \\
~ & ~ & ~ & ~ & r(x)3 \ar[r] & r(x)1 \\
} 
} 

and $<^\alpha_{PO}|y =$
\fbox{
\xymatrix{
~ & w(y)2 \ar@{|->}[dr] & ~ & ~ & ~ & ~ \\
~ & ~ & r(y)2 & ~ & ~ & ~ \\
} 
} 

are consistently linearizable.
$\alpha$ is also PRAM because

$<^\alpha_{PO}|(1,w) =$
\fbox{
\xymatrix{
w(x)1 \ar[r] & w(y)2 & ~ & ~ & ~ & ~ \\
~ & ~ &  & w(x)3  & ~ & ~ \\
} 
} 

$<^\alpha_{PO}|(2,w) =$
\fbox{
\xymatrix{
w(x)1 \ar[r] & w(y)2 \ar@{|->}[dr] & ~ & ~ & ~ & ~ \\
~ & ~ & r(y)2 \ar[r] & w(x)3 & ~ & ~ \\
} 
} 

$<^\alpha_{PO}|(3,w) =$
\fbox{
\xymatrix{
w(x)1 \ar[r] \ar@{|->}@/_3pc/[ddrrrrr] & w(y)2 & ~ & ~ & ~ & ~ \\
~ & ~ & ~ & w(x)3 \ar@{|->}[dr] & ~ & ~ \\
~ & ~ & ~ & ~ & r(x)3 \ar[r] & r(x)1 \\
} 
} 

have consistent linear extensions. But it is not Processor consistent
because the order of writes on $x$  in the, unique, linear extension of
$<^\alpha_{PO}|(2,w)$
is different from the order
in the linear extensions of $<^\alpha_{PO}|(3,w)$.

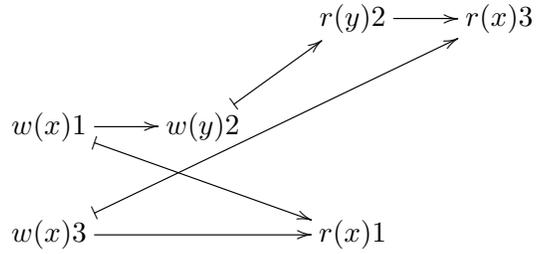
\begin{figure}[htbp]
\centerline{
\xymatrix{
~ & ~ & r(y)2 \ar[r] & r(x)3 \\
w(x)1 \ar@{|->}[drr] \ar[r] & w(y)2 \ar@{|->}[ur] & ~ & ~ \\
w(x)3 \ar@{|->}[uurrr] \ar[rr] & ~ & r(x)1 & ~ \\
} 
} 
\caption{Causal and Cache execution.}
\label{fig:causalAndCache}
\end{figure}

Finally, there is also the case of executions being Causal consistent
and Cache consistent but not Processor consistent.
Consider the execution $\alpha$ in figure \ref{fig:causalAndCache}.

$\alpha$ is Cache consistent because

$<^\alpha_{PO}|x =$
\fbox{
\xymatrix{
~ & ~ & ~ & r(x)3 \\
w(x)1 \ar@{|->}[drr] & ~ & ~ & ~ \\
w(x)3 \ar@{|->}[uurrr] \ar[rr] & ~ & r(x)1 & ~ \\
} 
} 

and
$<^\alpha_{PO}|y =$
\fbox{
\xymatrix{
~ & ~ & r(y)2 \\
~ & w(y)2 \ar@{|->}[ur] & ~ \\
} 
} 

have consistent linear extensions.

In addition, $\alpha$ is Causal because

$<^\alpha_{CR}|(1,w) =$
\fbox{
\xymatrix{
~ & ~ & r(y)2 \ar[r] & r(x)3 \\
w(x)1 \ar[r] & w(y)2 \ar@{|->}[ur] & ~ & ~ \\
w(x)3 \ar@{|->}[uurrr] & ~ & ~ & ~ \\
} 
} 

$<^\alpha_{CR}|(2,w) =$
\fbox{
\xymatrix{
w(x)1 \ar[r] & w(y)2 & ~ & ~ \\
w(x)3 & ~ & ~ & ~ \\
} 
} 

$<^\alpha_{CR}|(3,w) =$
\fbox{
\xymatrix{
w(x)1 \ar@{|->}[drr] \ar[r] & w(y)2 & ~ & ~ \\
w(x)3 \ar[rr] & ~ & r(x)1 & ~ \\
} 
} 

have consistent linear extensions.
However, all the consistent linear extensions of
$<^\alpha_{PO}|(1,w) (= <^\alpha_{CR}|(1,w))$
and
$<^\alpha_{PO}|(3,w) (= <^\alpha_{CR}|(3,w))$
have different orders for $w(2,x,1)$ and $w(3,x,3)$.

\subsection{Slow Consistency}

This model, \cite{Hutto:Slow:1990}, is a weaker version of both PRAM and Cache Consistency.

\begin{definition}{Slow Consistency}

$\alpha$ is an execution by a Slow memory $\equiv$

$$(\forall v \in {\cal V}, i \in {\cal P} \such : (<^\alpha_{PO} | (i, w(\cdot, v, \cdot) ) )
\ \textrm{is consistently linearizable} )$$

\fin
\end{definition}

\begin{figure}[htbp]
\centerline{
\xymatrix{
 w(x)1 \ar@/_3pc/@{|->}[drrrr] \ar[r] & w(x)2 \ar[r] & w(y)3 \ar[r] \ar@{|->}[dr] & r(y)4 & ~\\
  ~ & ~ & w(y)4 \ar[r] \ar@{|->}[ur] & r(y)3 \ar[r] & r(x)1 \\
} 
} 
\caption{Slow execution.}
\label{fig:slow}
\end{figure}
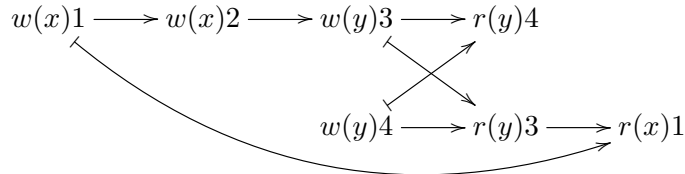

It is easy to see that figure
\ref{fig:causal} shows a Slow (and Causal and PRAM) execution that is not Cache.
On the  other hand,  the execution in  figure \ref{fig:cache}  is Slow
(and Cache) but not PRAM.
Combining them, see figure \ref{fig:slow}, we can build a Slow execution
not being Cache nor PRAM.
It is Slow because

\begin{itemize}

\item $<^\alpha_{PO}|(1,w(\cdot,x,\cdot) ) =$
\fbox{
\xymatrix{
 w(x)1  \ar[r] &  w(x)2 \ar[r] &  w(y)3 \ar[r] & r(y)4 & ~ \\
} 
} 
has this consistent
extension:
$w(1,x,1) ~ w(1,x,2) ~ w(1,y,3) ~ w(2,y,4) ~ r(1,y,4)$

\item $<^\alpha_{PO}|(2,w(\cdot,x,\cdot) ) =$
\fbox{
\xymatrix{
 w(x)1 \ar@{|->}[drrrr]  \ar[r] &  w(x)2  &  ~ & ~ & ~\\
  ~ & ~ & w(y)4 \ar[r] & r(y)3 \ar[r] & r(x)1 \\
} 
} 

has this consistent
extension:

$ w(2,y,4) ~ w(1,y,3) ~ r(2,y,3) ~ w(1,x,1) ~ r(2,x,1) ~ w(1,x,2) $

\item $<^\alpha_{PO}|(1,w(\cdot,y,\cdot) ) =$

\fbox{
\xymatrix{
 w(x)1 \ar[r] & w(x)2 \ar[r] & w(y)3 \ar[r]  & r(y)4 & ~\\
  ~ & ~ & w(y)4 \ar@{|->}[ur] & ~ & ~ \\
} 
} 

has this consistent extension:
$w(1,x,1) ~ w(1,x,2) ~ w(1,y,3) ~ w(2,y,4) ~ r(1,y,4)$

\item $<^\alpha_{PO}|(2,w(\cdot,y,\cdot) ) =$
\fbox{
\xymatrix{
  w(y)3  \ar@{|->}[dr] & ~ & ~\\
   w(y)4 \ar[r]  & r(y)3 \ar[r] & r(x)1 \\
} 
} 

has this consistent extension:
$ w(2,y,4) ~ w(1,y,3) ~ r(2,y,3) ~ w(1,x,1) ~ r(2,x,1) $

\end{itemize}

\subsection{Relations among Consistency Models}

Some indications  on how  consistency models  are related  appeared in
past  sections.    Here,  we  are  summarizing   them.   First,  table
\ref{tab:models} compiles the definitions of the consistency models.


\begin{table}[htbp]
\begin{tabular}{|r|p{9cm}|}
\hline
Model & $\alpha \in$ Model $\equiv$ \\
\hline 
\hline 
Sequential & $<^\alpha_{PO}$ is consistently linearizable \\
\hline
Causal & $(\forall i \in {\cal P} \such :\ <^\alpha_{CR} |(i,w)$ is
         consistently linearizable$)$ \\
\hline 
PRAM & $(\forall i \in {\cal P} \such :\ <^\alpha_{PO} |(i,w)$ is
         consistently linearizable$)$ \\
\hline 
Cache & $(\forall v \in {\cal V} \such :\ <^\alpha_{PO} |v$ is
         consistently linearizable$)$ \\
\hline 
Processor &
    $(\forall i, j \in {\cal P} \such : $

      ~~$<^\alpha_{PO}|(i,w), <^\alpha_{PO}|(j,w)$ have, respectively,

       ~~ consistent linear extensions $\beta_i, \beta_j$

       ~~ $\lAnd (\forall x \in {\cal V} \such : \beta_i|w(\cdot,x,\cdot)
          =\beta_j|(w(\cdot,x,\cdot) ) )$ \\

\hline
Slow & 
$(\forall v \in {\cal V}, i \in  {\cal P} \such : (<^\alpha_{PO} | (i,
w(\cdot, v, \cdot) ) ) $
is consistently linearizable $)$ \\
\hline      
\end{tabular}
\caption{Definition of Consistency Models}
\label{tab:models}
\end{table}

From these definitions, it immediately follows that
\begin{itemize}
\item Execution $\alpha$ is sequential $\implies$ $\alpha$ is a Causal,
  PRAM, Cache and Processor execution as well.
  This  is, a  Sequential memory  is also  a Causal,  PRAM, Cache  and
  Processor memory.
\item  Execution $\alpha$  is  Causal $\implies$  $\alpha$  is a  PRAM
  execution.
\item Execution $\alpha$ is Processor $\implies$ $\alpha$ is a PRAM
  and Cache execution.
\item Execution $\alpha$ is PRAM $\implies$ $\alpha$ is a Slow execution.
\item Execution $\alpha$ is Cache $\implies$ $\alpha$ is a Slow
  execution.\footnote{For this case, recall that
    \begin{itemize}
    \item $ <^\alpha_{PO} | (i, w(\cdot, v, \cdot) ) =
      (<^\alpha_{PO} | i ) \union
      (<^\alpha_{PO} | w(\cdot, v, \cdot) ) $
    \item The  reason for  an execution $\alpha$  being Cache  but not
      Slow, is not that $<^\alpha_{PO} | w(\cdot, v, \cdot)$ can't be
      linearized, or otherwise $<^\alpha_{PO}  | v$ would also
      fail to be, thus not being Cache either.
    \item But it is neither the case that there is a Cache execution
      which can't be extended  without respecting $<^\alpha_{PO} | i $.
      This  would imply  that $o_1(x)  <^\alpha_{PO} o_2(y)$  (with $x
      \neq y$)
      and that $o_2 (<^\alpha_{PO}|x \union \writes_\alpha)^* o_1$.
      This latter is not  possible, since neither  $<_{PO}$ nor
      $\writes$ can reach, by definition, actions before $o_2$.
    \end{itemize}
}

\end{itemize}

Now, we gather a set of executions in order to proof
some non-existent relationships among models.

\begin{enumerate}
\item Causal, PRAM, non-Sequential, non-Cache, non-Processor.
\label{causalSeq}

~
\xymatrix{
  w(x)1 \ar[r] \ar@{|->}[dr] & r(x)2 \\
  w(x)2 \ar[r] \ar@{|->}[ur] & r(x)1 
} 

\item PRAM, non-Sequential, non-Causal, non-Cache, non-Processor.
\label{PRAMSeq}

~
\xymatrix{
w(x)1 \ar@{|->}[dr] \ar@{|->}@/^1pc/[ddrrrr] & ~ & ~ & ~ & ~ \\
~ & r(x)1 \ar[r] & w(x)2 \ar@{|->}[dr]& ~ & ~ \\
~ & ~ & ~ & r(x)2 \ar[r] & r(x)1 \\
} 

\item Cache, non-Sequential, non-Causal, non-PRAM, non-Processor.
\label{cacheSeq}

~
\xymatrix{
 w(x)1 \ar@/_2pc/@{|->}[drrrr] \ar[r] & w(x)2 \ar[r] & w(y)3 \ar@{|->}[dr] & ~ & ~\\
  ~ & ~ & ~ & r(y)3 \ar[r] & r(x)1 \\
} 

\item Processor, non-Sequential, non-Causal, PRAM, Cache.
\label{procSeq}

~ 
\xymatrix{
w(x)1 \ar@{|->}[dr] \ar@{|->}@/^1pc/[ddrrrrr] & ~ & ~ & ~ & ~ & ~ \\
~ & r(x)1 \ar[r] & w(y)2 \ar@{|->}[dr]& ~ & ~ & ~ \\
~ & ~ & ~ & r(y)2 \ar[r] & w(x)3 \ar[r] & r(x)1 \\
} 

\item PRAM, Cache, non-Sequential, non-Causal, non-Processor
\label{pramCacheSeq}

 ~ 
\xymatrix{
w(x)1 \ar[r] \ar@{|->}@/_3pc/[ddrrrrr] & w(y)2 \ar@{|->}[dr] & ~ & ~ & ~ & ~ \\
~ & ~ & r(y)2 \ar[r] & w(x)3 \ar@{|->}[dr] & ~ & ~ \\
~ & ~ & ~ & ~ & r(x)3 \ar[r] & r(x)1 \\
} 

\item Causal, Cache, PRAM, non-Sequential, non-Processor.
\label{causalCacheSeq}

~
\xymatrix{
~ & ~ & r(y)2 \ar[r] & r(x)3 \\
w(x)1 \ar@{|->}[drr] \ar[r] & w(y)2 \ar@{|->}[ur] & ~ & ~ \\
w(x)3 \ar@{|->}[uurrr] \ar[rr] & ~ & r(x)1 & ~ \\
} 

\item   Slow,   non-PRAM,   non-Cache,   non-Causal,   non-Sequential,
  non-Processor.
\label{slow}

~
\xymatrix{
 w(x)1 \ar@/_3pc/@{|->}[drrrr] \ar[r] & w(x)2 \ar[r] & w(y)3 \ar[r] \ar@{|->}[dr] & r(y)4 & ~\\
  ~ & ~ & w(y)4 \ar[r] \ar@{|->}[ur] & r(y)3 \ar[r] & r(x)1 \\
} 
\end{enumerate}

Table \ref{tab:relationships} summarizes how memory models are related.
When a given memory model does not implies another different one,
the number  refers to the execution  in the previous list  that proves
it.

\begin{table}[htbp]
\begin{tabular}{|p{1.9cm}||c|c|c|c|c|p{1cm}|p{1.2cm}|}
\hline
Model x $\implies$  Model y & Seq. &  Causal & PRAM & Cache  & Proc. & Slow \\
\hline 
\hline 
Sequential & $\bullet$ & $\bullet$ & $\bullet$ & $\bullet$ & $\bullet$ & $\bullet$\\
\hline 
Causal & (\ref{causalSeq}) & $\bullet$ & $\bullet$ & (\ref{causalSeq}) & (\ref{causalSeq}) & $\bullet$ \\

\hline 
PRAM & (\ref{PRAMSeq}) & (\ref{PRAMSeq}) & $\bullet$ & (\ref{PRAMSeq}) & (\ref{PRAMSeq}) & $\bullet$ \\

\hline Cache & (\ref{cacheSeq})  & (\ref{cacheSeq}) & (\ref{cacheSeq}) & $\bullet$ & (\ref{cacheSeq}) & $\bullet$\\

\hline  Processor &  (\ref{procSeq}) &  (\ref{procSeq}) &  $\bullet$ & $\bullet$ & $\bullet$ & $\bullet$ \\

\hline  Slow &  (\ref{slow}) &  (\ref{slow}) &  (\ref{slow}) & (\ref{slow}) & (\ref{slow}) & $\bullet$ \\

\hline 
PRAM  $\lAnd$ Cache  & (\ref{pramCacheSeq})  & (\ref{pramCacheSeq})  & $\bullet$ & $\bullet$ & (\ref{pramCacheSeq}) & $\bullet$ \\

\hline 
Causal $\lAnd$ Cache &  (\ref{causalCacheSeq}) & $\bullet$ & $\bullet$ & $\bullet$ & (\ref{causalCacheSeq}) & $\bullet$\\

\hline 
Causal $\lAnd$ PRAM $\lAnd$ 
Cache $\lAnd$ Proc. 
&  ($\dagger$) &  $\bullet$  &  $\bullet$ &  $\bullet$  & $\bullet$  & $\bullet$ \\

\hline 
\end{tabular}
\caption{Model Relationships}
\label{tab:relationships}
\end{table}

We have found  an execution that is Causal, PRAM,  Cache and Processor
consistent at the same time, but it is not a Sequential execution.
This fact is explained in 
section \ref{sec:CORelation}.

\section{Relaxing Process Order}


So far, all models have been defined upon
the same notion of Process Order.
Now, we point out that Process Order can also be relaxed
without breaking its essence: {\em Lazy Process Order}.

Lazy  Process Order  is  a  subset of  Process  Order, where  original
relationships are preserved for a read and subsequent writes, and
among operations on the same variable.

\begin{definition}{Lazy Process Order}

$o_1 <_{LPO}^\alpha o_2 \equiv $

    \[
    o_1 <_{PO}^\alpha o_2 \lAnd \\
    \begin{cases}
      o_1 = r(\cdot, \cdot, \cdot) \\
      \lOr \\
      o_1 = \cdot(\cdot, x, \cdot) \lAnd
      o_2 = \cdot(\cdot, x, \cdot) 
    \end{cases}
    \]

\fin
\end{definition}

In order to justify Lazy Program Order, consider this execution:
\centerline{
\xymatrix{
w(x)1 \ar[r] & w(x)2 \ar@{~>}[dr] \\
~ & ~ & w(y) \ar[r] & r(x)?
} 
}

where a memory system ensures that the global order of writes is
$w(1,x,1) w(1,x,2) w(2,y,\cdot)$.
At a given  moment, the value $1$  is available at process  2, but 
value $2$ is not yet ready.  If Process Order must be respected, the
read have to wait until value  2 is available. But $w(2,y,\cdot)$ and
$r(2,x,?)$  are not  related under  Lazy Process  Order. Hence,  it is
allowed for the read to get value 1.

A  case  of this  situation  is  also  the  Cache execution  shown  in
figure \ref{fig:cache}
in page \pageref{fig:cache}.

\section{Synchronized Consistency Models}

In addition  to write and read  accesses, a memory system  can support
synchronization primitives in order to simplify the coordination among
the processes communicating through it.   These primitives can be used
to set up dependencies among the  rest of memory accesses, and to mark
points  where the  state  of the  memory  is made  common  to all  the
processes.

In order to define these type of models we have to introduce
a  set of  synchronization  variables ${\cal  S}$,  and the  following
operations.

\begin{definition}{Acquire and Release}

\begin{itemize}
\item $rel(i,s)$ to
  denote that process $i \in \cal P$ {\em releases} 
  the variable $s \in \cal S$.
\item $acq(i,s)$ to
  denote that process $i \in \cal P$ {\em acquires} the
  variable $s \in \cal S$.
\end{itemize}

\fin
\end{definition}

Now, the definition  of {\em valid execution} is extended  to include the
fact that these primitives guarantee mutual exclusion.

\begin{definition}{Valid Execution}

  $\alpha$ is a valid execution $\equiv$

  ~~~ $\alpha | (w, r) $ is valid according to definition \ref{def:ValidExecution} $\lAnd$
  for $\beta =  \alpha | (acq, rel)$ and  any synchronization variable
  $s$
$$
  \begin{cases}
      \beta |  s  = \epsilon \\
      \lOr \\
      \beta | s  = 
      acq(i,s) rel(i,s) \beta' \lAnd  \beta' \textrm{\ is a valid execution\ }
  \end{cases}
$$

\fin
\end{definition}

As of  the previous definition,  we define the {\em  mutual exclusion}
order for a valid execution.

\begin{definition}{Mutual Exclusion Order}

  $a <_{ME}^\alpha b \equiv$

  $$
  (\exists s \in \textrm{\cal S} \such : \alpha | s = \alpha_1 a \alpha_2 b \alpha_3
  \lAnd \alpha \textrm{\ is valid.})
  $$

\fin
\end{definition}

The relationship between a release operation and the corresponding
subsequent  acquire  gaining  the  mutual  exclusion  right  over  the
synchronization variable can be defined like a 
read-write analogy for synchronizations.

\begin{definition}{Writes-to for Synchronizations}

$rel(i,s) \writes_\alpha acq(j,s)$ $\equiv$

$$
\alpha | ( acq(\cdot,s), rel(\cdot,s) ) =
\alpha_1 rel(i,s) acq(j,s) \alpha_2
$$

\fin
\end{definition}

\begin{definition}{$D_-$, $D_+$, $D$ and $<_{SO}^\alpha$}

Every synchronization action $o$ has  two associated sets $D_-(o)$ and
$D_+(o)$  that group,  respectively, the  ordinary accesses  intended
to precede and follow $o$.

The Synchronization Order (SO) formalizes this idea on how synchronizations induce the
ordering of the rest of actions.

  $D(s)$  is  the   set  of  ordinary  accesses   {\em  dependent}  on
  synchronization variable $s$.

$a <_{SO}^\alpha b$ $\equiv$

$$
  \begin{cases}
    a <_{ME}^\alpha b \\
    \lOr \\
    a \in D_-(b) \\
    \lOr \\
    b \in D_+(a) \\
    \lOr \\
    (\exists c \in \alpha \such : a <_{ME}^\alpha c <_{ME}^\alpha b)
  \end{cases}
$$

\fin
\end{definition}

The minimum consistency that synchronized models support is Slow
consistency. This is, between  two consecutive synchronizations,
the order among writes on the same variable by the same process will be respected.

\begin{definition}{Synchronized Consistency Model}

$\alpha$ is an execution by a Synchronized memory $\equiv$

$$(\forall v \in {\cal V}, i \in {\cal P} \such : <_{SO}^\alpha \union
  (<^\alpha_{PO} | (i, w(\cdot, v, \cdot) ) ) \ \textrm{is consistently linearizable} )$$

Different Synchronized Models are  defined and distinguished by means
of $D_-$ and $D_+$.

\fin
\end{definition}

The following  table gathers the  definitions of the  main Synchronized
Models:         Weak        \cite{Dubois:Memory:1986},         Release
\cite{Gharachorloo:Memory:1990}, and Entry \cite{Bershad:Midway:1993}.

\begin{tabular}{|p{2cm}||l|}
\hline
{\bf Model} & {\bf Definition} \\
\hline
Weak & $D_-(o) = \{ e : e <_{PO}^\alpha o \}$  \\
~ &  $D_+(o) = \{ e : o <_{PO}^\alpha e \}$ \\

\hline 

Release   &  $D_-(rel(\cdot,\cdot))   =   \{  e   :  e   <_{PO}^\alpha rel(\cdot,\cdot) \}$ \\
~ & $D_+(acq(\cdot,\cdot)) = \{ e : acq(\cdot,\cdot) <_{PO}^\alpha e \}$ \\

\hline

Lazy  & $D_-(acq(\cdot,\cdot)) = \{ e :
(\exists  rel   \such ~ rel   \writes_\alpha  acq(\cdot,  \cdot) )  :  e
<_{PO}^\alpha rel  \}$ \\
Release & $D_+(acq(\cdot,\cdot)) = \{ e : acq(\cdot,\cdot) <_{PO}^\alpha e \}$ \\

\hline

Entry & $D_-(acq(\cdot,s)) = \{ e :
(\exists  rel   \such ~ rel(\cdot,s)   \writes_\alpha  acq(\cdot,  s) )  :  e
<_{PO}^\alpha rel$ \\
~ & $~~~~~~~~ \lAnd e \in D(s) \}$ \\

~ & $D_+(acq(\cdot,s) = \{ e : acq(\cdot,s) <_{PO}^\alpha e
\lAnd e \in D(s) \}$ \\

\hline

\end{tabular}

\newpage
\appendix

\section{When an Execution has a Consistent Linear Extension}
\label{sec:CORelation}

As  we  know,  a  consistent  linear extension  of  a  given  relation
$\related_\alpha$ may not, by
definition,
have read operations fetching an overwritten value.  Therefore,
whenever an execution  contains a related write-read  pair, $w \writes
r$,  and a  different write  $w'$,  all on  the same  variable; it  is
necessary that  $w'$ is either  before or after  $w \writes r$  in any
linear  extension.  In  addition,  linear extensions  do respect
some other relation $\related$ as well; usually Process Order or Causal
relations, maybe restricted.

The point is: how $w \writes  r$ is related to $w'$ in $\related$?.
If $w' \related w$, there is  no need to introduce any new dependency;
as well  as it  is the case  for $r \related  w'$, because  either $w'
\related w  \writes r$  or $w  \writes r  \related w'$  alredy defines
their relative order in a linear extension.

But for the cases
$w \related w'$ and
$w' \related r$ we can infer new dependencies to be held
in every {\bf co}nsistent linear extension.

\begin{definition}{co extension of $\related_\alpha$}
\begin{itemize}

\item WW dependency

$w' \to_{co}^{\related_\alpha} w \equiv w \writes_\alpha r \lAnd w' \related_\alpha r$

\item RW dependency

$r \to_{co}^{\related_\alpha} w' \equiv w  \writes_\alpha r \lAnd w \related_\alpha w'$

\item co extension of $\related_\alpha$

  $co(\related_\alpha) = (\related_\alpha \cup \to_{co}^{\related_\alpha})^*$

\end{itemize}

\fin
\end{definition}

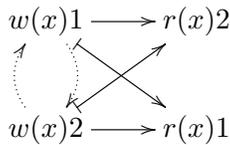
\begin{figure}[htbp]
\centerline{
\xymatrix{
  w(x)1 \ar[r] \ar@{|->}[dr] \ar@/^1pc/@{.>}[d] & r(x)2 \\
  \ar@/^1pc/@{.>}[u] w(x)2 \ar[r] \ar@{|->}[ur] & r(x)1 
} 
} 
\caption{Causal execution with CO dependencies.}
\label{fig:causalWithCO}
\end{figure}

Let's show some examples of 
executions augmented with $<_{CO}$ dependencies.
In figure \ref{fig:causalWithCO}, we can see a causal execution with WW
dependencies added:
$w(1,x,1) \writes r(2,x,1) \lAnd w(2,x,2) \related r(2,x,1)$
and
$w(2,x,2) \writes r(1,x,2) \lAnd w(1,x,1) \related r(1,x,2)$.
This execution is not sequentially consistent.
CO dependencies have  created a cycle in the graph  that just shows us
there is no consistent linear extension for it.

\begin{figure}[htbp]
\centerline{
\xymatrix{
 w(x)1 \ar@/_2pc/@{|->}[drrrr] \ar[r] & w(x)2 \ar[r] & w(y)3 \ar@{|->}[dr] & ~ & ~\\
  ~ & ~ & ~ & r(y)3 \ar[r] & r(x)1 \ar@/_2pc/@{.>}[ulll] \\
} 
} 
\caption{Cache execution with CO dependencies.}
\label{fig:cacheWithCO}
\end{figure}
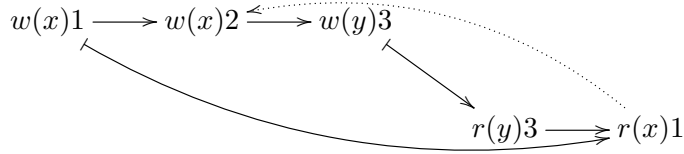

Just  one more  example.  Figure  \ref{fig:cacheWithCO} shows  a Cache
execution with just a RW dependency added:
$w(1,x,1) \writes r(2,x,1) \lAnd w(1,x,1) \related r(1,x,2)$.
Again,  this execution is
not sequentially consistent and the RW arrow makes the graph cyclic.

By  applying $co(\related_\alpha)$  once,  we obtain  a new  $\related'$
binary  relation  which   can  be  used  again  to   discover  new  CO
dependencies.
Hence,
let's generalize the $co$ operator.


\begin{definition}{$CO(\related_\alpha)$}

$$CO(\related_\alpha) = \lim_{n\to\infty}
\underbrace{co( \cdots co}_n(\related_\alpha) )
$$

\fin
\end{definition}

When a new CO  dependency is added, new ones could  be found when $co$
is applied again.
But because executions are finite, the limit exists.

It  easily   follows  from  its   definition  that  the   acyclicity  of
$CO(\related_\alpha)$ is  a necessary condition for  $\related_\alpha$ to
have  a consistent  linear extension.   The reverse  statement is  not
true.  There  are non-consistently linearizable executions  with a
cycle-free $CO$ extension.

\begin{figure}[htbp]

\centerline{\epsfig{file=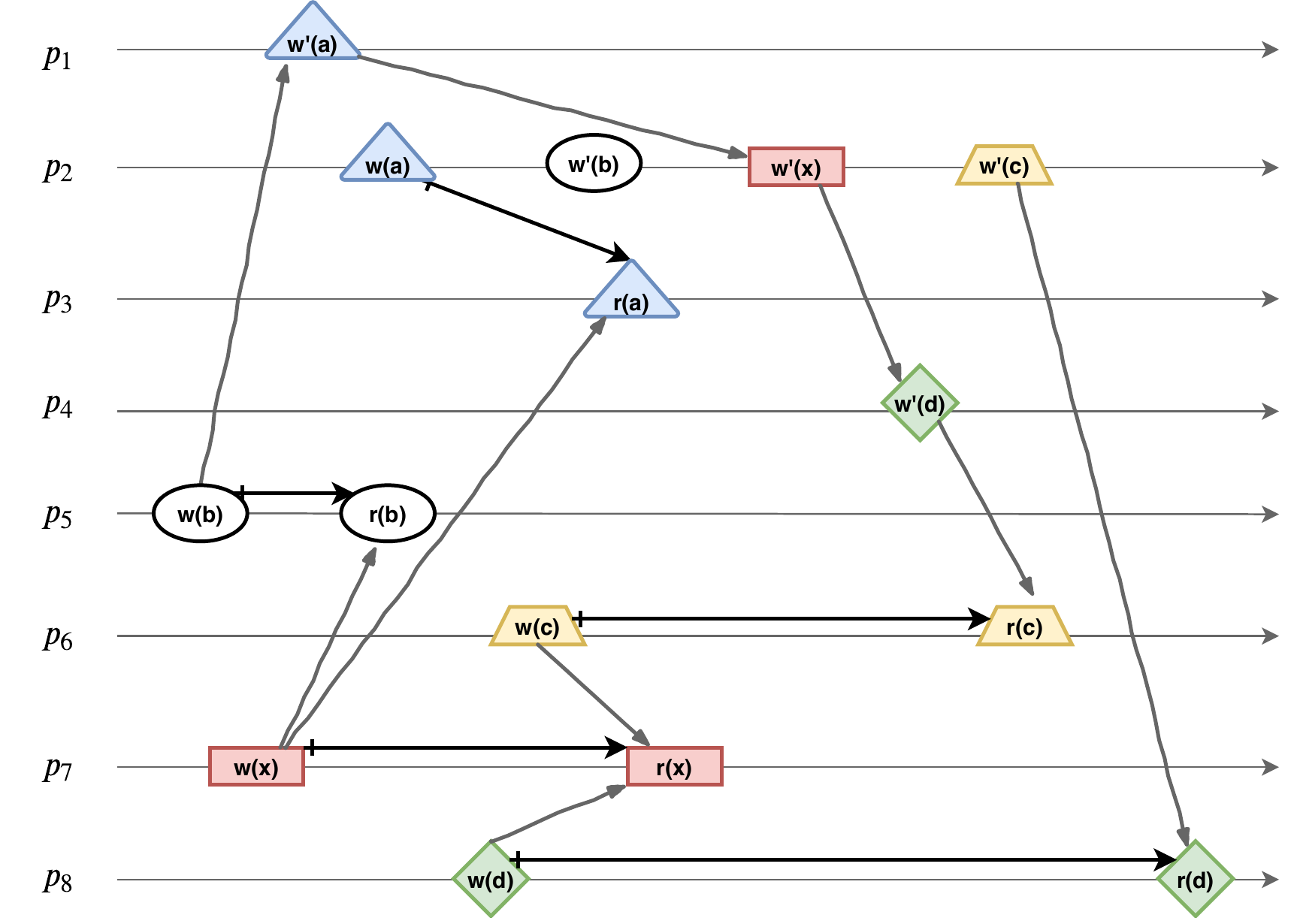,width=\linewidth}}

\caption{Non-Sequential execution with acyclic CO.}
\label{fig:superexecution}
\end{figure}

This can happen in executions  with {\em unrelated $w~r~w'$ triplets}:
operations  on the  same  variable with  $w \writes  r$  and $w'$  not
related either  to $w$ or  $r$.  For  some executions, it  could occur
that  any choice,  $w' \related  w$ or  $r \related  w'$, to  obtain a
linear  extension  leads  to  an overwritten  value  for  a  different
variable.      Consider     the    diagram-execution     in     figure
\ref{fig:superexecution} with horizontal lines  for process order, and
$\writes$ arrows for writes-to relationship as usual. Each one of the rest
of arrows indicate a causal dependency, induced by a chain of write-to
(on different  variables) and  process order actions.   The particular
actions of the  chain are not shown  to keep the diagram  as simple as
possible.

The  actions shown  in figure  \ref{fig:superexecution} are  unrelated
triplets  on different  variables,  easily recognizable  by shape  and
color.  For each  one of them, there  is no path between  $w'$ and the
pair $w$ and $r$ ($w \writes r$).

Let's  consider the  triplet  on  variable $x$.  Because  there is  no
imposed
WW nor RW dependency, in order to have a consistent linear
extension:

\begin{itemize}
\item If we add a WW dependency $w(x)' \related w(x)$ (WW1), it appears the path
  $w'(a) ~  w'(x) ~ w(x) ~  r(a)$ which forces a  WW dependency $w'(a)
  \related w(a)$ (WW2).
  But now, the following path exists:
$w(b) ~ w'(a)' ~ w(a) ~ w'(b) ~ w'(x) ~ w(x) ~ r(b)$
with $w'(b)$ between  $w(b) \writes r(b)$ and  therefore preventing it
from being linearizable. The following figure shows this case:

\centerline{\epsfig{file=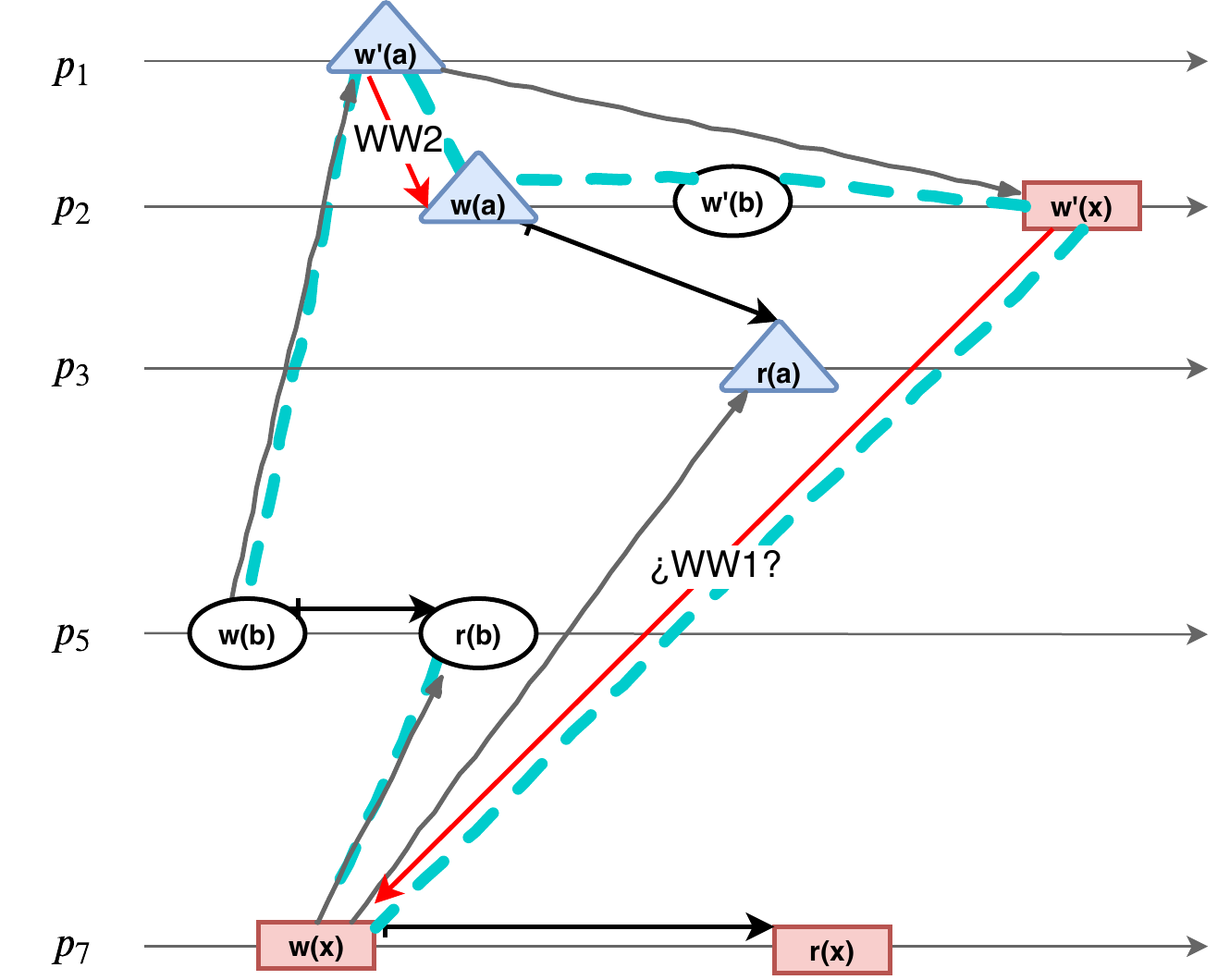,width=0.8\linewidth}}

\item If we add a RW dependency $r(x) \related w'(x)$ (RW1), then, $w(d) ~
  r(x) ~  w'(x) ~ w'(d)$  implies a  new RW dependency  $r(d) \related
  w'(d)$ (RW2). Now, there exists the path
  $ w(c) ~ r(x) ~ w'(x) ~ w'(c) ~ r(d) ~ w'(d) ~ r(c) $
  where $w'(c)$ is between $w(c) \writes r(c)$.
  This case is represented here:

  ~

\centerline{\epsfig{file=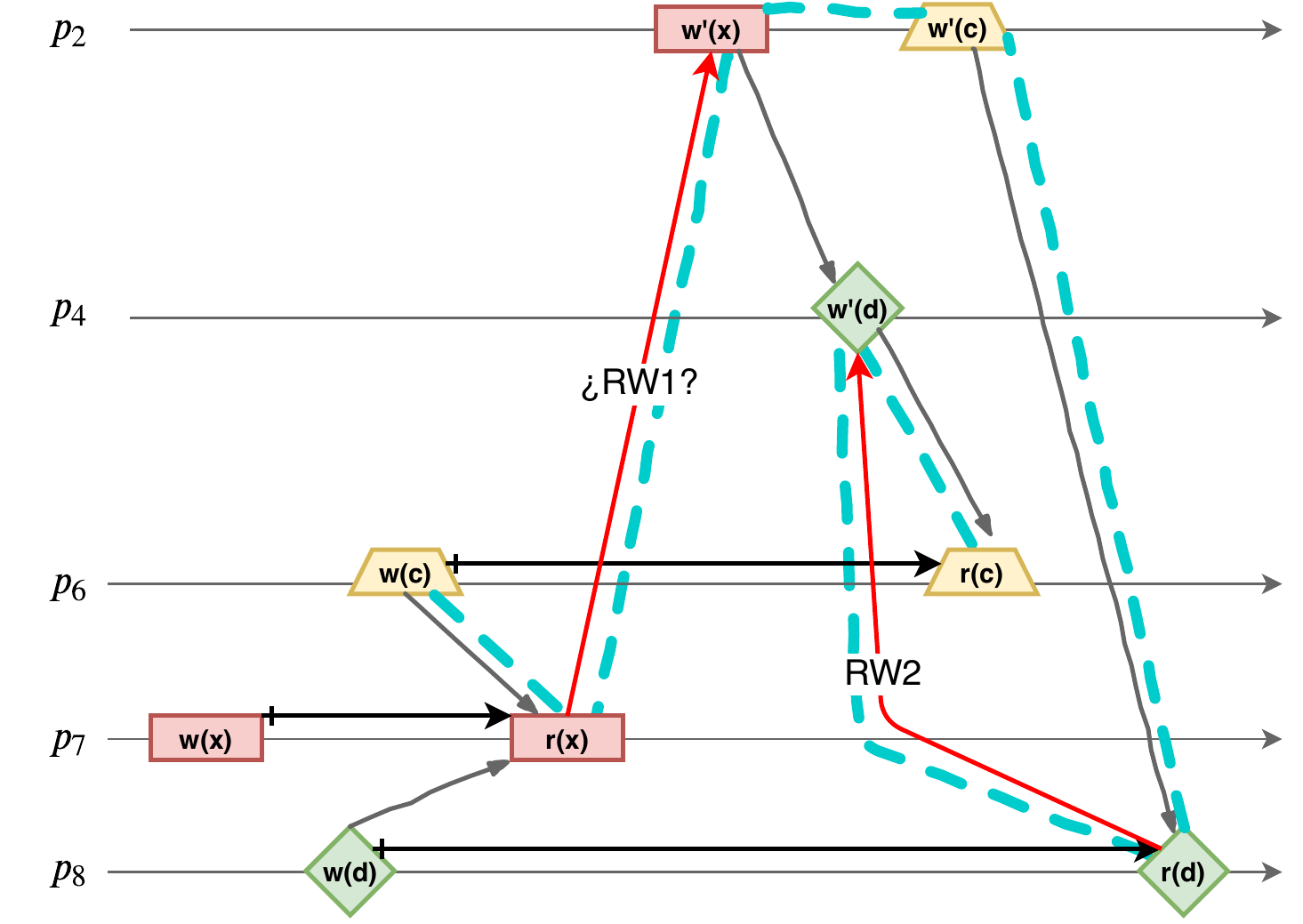,width=0.8\linewidth}}

\end{itemize}

In sum,  the execution  in figure  \ref{fig:superexecution} is  a case
where  the CO  extension  of  $<_{PO}$ is  acyclic  but the  execution
does not have a consistent linear extension.

Finally,

\begin{itemize}

\item Because each process in the execution of figure \ref{fig:superexecution} 
has at  most one read  $r$ and  $w$ ($w \writes  r$) and $w'$  are not
causally related, it is
clear that the execution is Causal.

\item The following sequence:

$$
w(b)
~
r(b)
~
w'(a)
~
w(a)
~
r(a)
~
w'(b)
~
w'(x)
~
w(x)
~
r(x)
~
w(d)
~
r(d)
~
w(c)
~
r(c)
~
w'(d)
~
w'(c)
$$

is a  consistent extension of $<_{PO}^\alpha|(i,w)$,  for all processes
simultaneously, which proves the execution in figure \ref{fig:superexecution} 
is Processor.

\end{itemize}


%


%
%

%

%


\newpage
\bibliographystyle{plain}
\bibliography{texSections/bibliography}

\end{document}